\documentclass[12pt]{elsarticle}%
\usepackage{lineno,hyperref}
\usepackage{epsfig,amssymb,amsmath,version,amsthm}
\usepackage{amssymb,version,graphicx,fancybox,mathrsfs,multirow}
\usepackage{color}
\usepackage{accents}
\usepackage{appendix}
\usepackage{algorithm}
\usepackage{algorithmicx}
\usepackage{algpseudocode}
\usepackage{amsmath}
\usepackage{amsfonts}
\usepackage{amssymb}
\usepackage{graphicx}%
\RequirePackage{fix-cm}
\newtheorem{thm}{Theorem}

\newtheorem{Rem}{Remark}

\DeclareMathOperator{\supp}{supp}

\renewcommand{\d}{\ensuremath{\,\mathrm{d}}}

\renewcommand{\vec}[1]{\ensuremath{\boldsymbol{#1}}}

\renewcommand{\phi}{\varphi}
\renewcommand{\epsilon}{\varepsilon}

\begin{document}
\begin{frontmatter}
	\title{PhaseDNN - A Parallel Phase Shift Deep Neural Network for Adaptive Wideband Learning}
	

\address[add1]{Department of Mathematics, Southern Methodist University, Dallas, TX 75275, USA.}
	\address[add2]{LCSM, Ministry of Education, School of Mathematics and Statistics, Hunan Normal University, Changsha, Hunan 410081, P. R. China}

\author[add1]{Wei Cai}
	\author[add2,add1]{Xiaoguang Li}
    \author[add1]{Lizuo Liu}

	\begin{abstract}
		In this paper, we propose a phase shift deep neural network (PhaseDNN) which provides a wideband convergence in approximating a high dimensional function during its training of the network. The PhaseDNN utilizes the fact that many DNN achieves convergence in the low frequency range first, thus, a series of moderately-sized of DNNs are constructed and trained in parallel for ranges of higher frequencies.  With the help of phase shifts in the frequency domain, implemented through a simple phase factor multiplication on the training data,  each DNN in the series will be trained to approximate the target function's higher frequency content over a specific range. Due to the phase shift, each DNN achieves the speed of convergence as in the low frequency range. As a result, the proposed PhaseDNN system is able to convert wideband frequency learning to low frequency learning, thus allowing a uniform learning to wideband high dimensional functions with frequency adaptive training. Numerical results have demonstrated the capability of PhaseDNN in learning information of a target function from low to high frequency uniformly.
	\end{abstract}
	
	
\end{frontmatter}

\section{Introduction}

Deep neural network (DNNs) have shown great potential in approximating high
dimensional functions without suffering the curse of dimensionality of
traditional approximations based on Lagrangian interpolation or spectral
methods. Recently, it has been shown in \cite{xu18, xu19} that some common
NNs, including fully connected and CNN with tanh and ReLU activation
functions, demonstrated a frequency dependent convergence behavior, namely,
the DNN during the training are able to approximate the low frequency
components of the targeted functions first before the higher frequency
components, named as F-Principle of DNNs \cite{xu18}. The stalling of DNN
convergence in the later stage of training could be mostly coming from
learning the high frequency component. This F-principle behavior of DNN is
just the opposite to the convergence behavior of the traditional multigrid
method (MGM) \cite{bandt77} for approximating the solutions of PDEs where the
convergence occurs first in the higher frequency end of the spectrum, due to
the smoothing operator employed in the MGM. The MGM takes advantage of this
fast high frequency error reduction in the smoothing iteration cycles and
restrict the original solution of a fine grid to a coarse grid, then
continuing the smoothing iteration on the coarse grid to reduce the higher end
frequency spectrum in the context of the coarse grid. This downward
restriction can be continued until errors over all frequency are reduced by a
small number of iteration on each level of the coarse grids.

In this work, we propose a wideband DNN in error reductions in the
approximation for all frequencies of the targeted function by making use of
the faster convergence in the low frequencies of the DNN during its training.
We first will train a neural network $T_{\star}(\vec{x},\theta^{(n_{0})})$
until it will achieve a sufficient error reduction for the frequency range
$|k|<K_{0}$ for a prescribed $K_{0}$ within $n_{0}$ training epoches. In order
to further learn the high frequency component of the target function, we
decompose the residual, i.e. the difference between the target function and
$T_{\star}(\vec{x},\theta^{(n_{0})}),$ into a series of functions with
frequency spectrum through a partition of the unit (POU) in the $k$-space. To
learn each of these function of specific frequency range, we will employ a
phase shift in the $k$-space to translate its frequency to the range
$|k|<K_{0}$, the phase shifted function now only has low frequency content and
can be learned by common DNN with a small number of epoches of training. This
series of DNNs with phase shifts will be named Phase-shift deep Neural Network (PhaseDNN).

It should be noted that during the construction of PhaseDNN, only the original
set of training data will be needed, but it will be modified to account for
the frequency shifts, implemented simply by multiplying the training data by
an phase factor $e^{\pm\omega_{j}x}$ on the residuals from the previous neural
network approximations. We also allude the way PhaseDNN achieve approximation
over increasing range of frequency with each additional network to that of
wavelet approximations, which generate better and better resolution of
approximation by adding wavelet subspaces produced by dialation of mother
wavelet to account for higher frequency \cite{dau92}\cite{cai96}. In the case
of PhaseDNN though phase shifts are used to cover higher frequencies. The
substraction between the target function provided by the training data and the
lower frequency network approximation is similar to the strategy employed in
construction of multi-resolution interpolation for PDE solutions\cite{cai96}.

The rest of the paper will be organized as follows. In section 2, we will
review the fast low frequency convergence property of neural network with tanh
activation function. Section 3 will present the new phase shift deep neural
network.  Finally, a conclusion and discussions will be given in Section 4.

\section{Deep Neural Network and frequency dependent training of DNN}

A deep neuron network (DNN) is a sequential alternative composition of linear
functions and nonlinear activation functions. Given $m,n\geqslant1$, let
$\Theta(\vec{x}):\mathbb{R}^{n}\rightarrow\mathbb{R}^{m}$ is a linear function
with the form $\Theta(\vec{x})=\vec{W}\vec{x}+\vec{b}$, where $\vec{W}%
=(w_{ij})\in\mathbb{R}^{n\times m}$, $\vec{b}\in\mathbb{R}^{m}$. The nonlinear
activation function $\sigma(u):\mathbb{R}\rightarrow\mathbb{R}$. By applying
this function componentwisely, we can extend activation function to
$\sigma(u):\mathbb{R}^{n}\rightarrow\mathbb{R}^{n}$ naturally. A DNN with
$L+1$ layers can be expressed as a compact form
\begin{equation}
\begin{aligned} T(\vec{x}) &= T^L(\vec{x}),\\ T^l(\vec{x}) & = [\Theta^l\circ\sigma](T^{l-1}(\vec{x})), \quad l=1,2,\dots L, \end{aligned} \label{eq:dnniter}%
\end{equation}
with $T^{0}(\vec{x})=\Theta^{0}(\vec{x})$. Equivalently, we can also express
it explicitly:
\begin{equation}
T(\vec{x})=\Theta^{L}\circ\sigma\circ\Theta^{L-1}\circ\sigma\cdots\circ
\Theta^{1}\circ\sigma\circ\Theta^{0}(\vec{x}). \label{eq:dnn}%
\end{equation}
Here, $\Theta^{l}(\vec{x})=\vec{W}^{l}\vec{x}+\vec{b}^{l}:\mathbb{R}^{n^{l}%
}\rightarrow\mathbb{R}^{n^{l+1}}$ are linear functions. This DNN is also said
to have $L$ hidden layers. The $l$-th layer has $n^{l}$ neurons.

In the following text, we only consider DNN with one dimensional input layer
and one dimensional output layer. Namely, $\Theta^{0}(x):\mathbb{R}%
\to\mathbb{R}^{n^{1}}$, $\Theta^{L}(\vec{x}):\mathbb{R}^{n^{L-1}}\to
\mathbb{R}$. That is, $T(\vec{x}):\mathbb{R}\to\mathbb{R}$. The activation
function is chosen to be $\sigma(u)=\tanh(u)$.

In this paper, we consider the problem of approximating a function by DNN
through training. Given a function $f(x)\in\mathbb{L}^{1}(\mathbb{R}%
)\cap\mathbb{L}^{2}(\mathbb{R})$ (This choice is only to ensure the existence
of Fourier transform and loss function for analysis reason.), we are going to
minimize
\begin{equation}
L(\vec{W}^{0},\vec{b}^{1},\vec{W}^{1},\vec{b}^{1},\dots,\vec{W}^{L},\vec
{b}^{L})=\left\Vert {f(\vec{x})-T(\vec{x})}\right\Vert _{2}^{2}=\int_{-\infty
}^{+\infty}\left\vert f(\vec{x})-T(\vec{x})\right\vert ^{2}x. \label{eq:loss}%
\end{equation}
We consider this problem in frequency space. We define Fourier transform and its inverse of a
function $f(\vec{x})$ by
\begin{equation}\label{eq:fourier}%
\mathcal{F}[f](k)=\frac{1}{\sqrt{2\pi}}\int_{-\infty}^{+\infty}f(x)e^{-ikx}\d x,\quad\quad \mathcal{F}^{-1}[\widehat{f}](x) =  \frac{1}{\sqrt{2\pi}}\int_{-\infty}^{+\infty}\hat{f}(k)e^{ikx}\d k.
\end{equation}
Let $D(k)=\mathcal{F}[T-f](k)=A(k)e^{i\phi(k)}$, $L(k)=\left\vert
D(k)\right\vert ^{2}$. By Paseval's equality,
\begin{equation}
L(\vec{W}^{0},\vec{b}^{1},\vec{W}^{1},\vec{b}^{1},\dots,\vec{W}^{L},\vec
{b}^{L})=\int_{-\infty}^{+\infty}L(k)k. \label{eq:lossfourier}%
\end{equation}

Let $\theta$ denotes all the parameters in DNN. Namely,
\[
\theta=(\vec{W}_{1}^{0},\vec{W}_{2}^{0}\dots,\vec{W}_{n^{1}}^{0},\vec{b}%
^{0},\vec{W}_{11}^{1},\dots,\vec{W}_{n^{1} n^{2}}^{1},\vec{b}_{1}^{1}\dots
\vec{b}^{1}_{n^{2}}\dots)\in\mathbb{R}^{p}.
\]
Here, $p=n^{1}+1+(n^{1}+1)\times n^{2} + (n^{2}+1)\times n^{3}+\dots(n^{L}+1)$
is the number of the parameters. The F-principle states the relative changing
rate of $\frac{\partial L(k)}{\partial\theta_{j}}$ for different frequency
$k$. We cite the following result from \cite{xu19}

\begin{thm}
Given a function $f(\vec{x})\in\mathbb{L}^{1}(\mathbb{R})\cap\mathbb{L}%
^{2}(\mathbb{R})$ and a DNN function $T(\vec{x})$ defined by
\eqref{eq:dnniter}, we scale $T(\vec{x})$ by $\Theta^{0}(\vec{x})=\epsilon
\vec{W}^{0}\vec{x}+\vec{b}$, where $\epsilon>0$ is a small parameter. Suppose
$\vec{W}_{rs}^{l}\neq0$ for all $0\leqslant l\leqslant L-1$, $1\leqslant
r\leqslant n^{l}$ and $1\leqslant s\leqslant n^{l+1}$. For any frequency
$k_{1}$ and $k_{2}$ such that $\left|  \mathcal{F}[f](k_{1}) \right|  >0$,
$\left|  \mathcal{F}[f](k_{2}) \right|  >0$ and $\left|  k_{2} \right|
>\left|  k_{1} \right|  >0$, there exists a $\delta>0$ such that when
$\epsilon<\delta$,
\[
\frac{\mu\left(  \{\theta|\left|  \frac{\partial L(k_{1})}{\partial\theta_{j}}
\right|  >\left|  \frac{\partial L(k_{2})}{\partial\theta_{j}} \right|  \}\cap
B_{\delta}\right)  }{\mu(B_{\delta})}>1-C\exp(-c/\delta)
\]
holds for all $j$, where $B_{\delta}=\{\vec{x}\in\mathbb{R}^{p}|\left|
\vec{x} \right|  <\delta\}$, $\mu(\cdot)$ is the Lebesgue measure and $c>0$,
$C>0$ are two constants.
\end{thm}


\section{Parallel Phase shift DNN (PhaseDNN)}

The F-principle and estimation of $L(k)$ in Theorem 1 suggest that when we
apply a gradient-based training method to loss function, the low frequency
part of loss function converges faster than the high frequency part. So when
we train a DNN $T_{\star}(x)$ to approximate $f(x)$, there exists positive
frequency number $K_{0}$ and an integer $n_{0}$ such that after $n_{0}$ steps
of training, the Fourier transform of training function $\mathcal{F}[T_{\star
}](\vec{k};\theta^{(n_{0})})$ should be a good approximation of $\mathcal{F}%
[f](\vec{k})$ for $\left\vert k\right\vert <K_{0}$, where $\theta^{(n_{0})}$
are parameters obtained after $n_{0}$ steps of training.

\begin{itemize}
\item Frequency selection kernel $\phi_{j}^{\vee}(x)$
\end{itemize}

In order to speed up the learning of the higher frequency content of the
target function $f(x)$, we will employ a phase shift technique to translate
higher frequency spectrum $\widehat{f}(k)$ to the frequency range of
$[-K_{0},K_{0}]$. Such a shift in frequency is a simple phase factor
multiplication on the training data in the physical space, which can be easily implemented.

For a given $\Delta k$, let us assume that
\[
\supp\widehat{f}(k)\subset\lbrack-m\Delta k,m\Delta k].
\]
Before we can carry out this scheme, we construct a mesh for the interval
$[-m\Delta k,m\Delta k]$ by
\begin{equation}
\omega_{j}= j\Delta k,j=-m,\cdots,m,\label{mesh}%
\end{equation}

Then, we introduce a POU for the domain $[-M,M]$ associated with the mesh as%

\begin{equation}
1=%
{\displaystyle\sum\limits_{j=-m}^{m-1}}
\phi_{j}(k),\text{ }k\in\lbrack-M,M].\label{pou}%
\end{equation}
The simplest choice of $\phi_j(k)$ is
\begin{equation}\label{eq:phij}
  \phi_j(k)=\phi(\frac{k-\omega_{j}}{\Delta k}),
\end{equation}
where $\phi(k)$ is just the characteristic function
of $[-1,1],$ i.e.,%

\[
\phi(k)=\chi_{\lbrack0,1]}(k).
\]
The inverse Fourier transform of $\phi(k)$, indicated by $\vee,$ is
\begin{equation}\label{eq:invfphi}
  \phi^{\vee}(x) = \frac{1}{\sqrt{2\pi}}e^{i\frac{x}{2}}\frac{\sin x}{x}.
\end{equation}

For a smoother function $\phi(k)$ whose inverse Fourier transform, which will
act as a frequency selection kernel in $x$-space, has a faster decay condition,
we could use B-Splines $B_{m}(k),$defined by the following recursive
convolutions. Namely,%

\[
B_{1}(k)=\chi_{\lbrack0,1]}(k),
\]

\begin{equation}
B_{m}(k)=B_{m-1}(k)\ast\chi_{\lbrack0,1]}(k),m=2,\cdots.\label{b-spline}%
\end{equation}
It can be easily shown that%

\[
\supp B_{m}(k)=[0,m],
\]
and the inverse Fourier transform of $B_{m}(k)$ is%

\begin{align*}
B_{m}^{\vee}(x)  &  =\left(  B_{1}^{\vee}(x)\right)  ^{m}=\left(  \frac
{1}{\sqrt{2\pi}}e^{i\frac{x}{2}}\frac{\sin\frac{x}{2}}{\frac{x}{2}}\right)
^{m}\\
&  =\frac{e^{i\frac{mx}{2}}}{\left(  2\pi\right)  ^{m/2}}\left(  \frac
{\sin\frac{x}{2}}{\frac{x}{2}}\right)  ^{m}.
\end{align*}

In most cases, we will just use the cubic B-Spine. We choose
\begin{equation}
\phi(k)=B_{4}(k+2),\label{cubicspline}%
\end{equation}
and $\phi_j(k)=\phi(k/\Delta k-j)$. Therefore,
\[
\phi^{\vee}(x)=\frac{1}{\left(  2\pi\right)
}\left(  \frac{\sin\frac{x}{2}}{\frac{x}{2}}\right)  ^{4}=O(\frac{1}{|x|^{4}%
})\text{ as x}\rightarrow\infty,
\]
which defines the \emph{frequency selection kernel}%

\[
\phi_{j}^{\vee}(x)=e^{-ij\Delta kx}\phi^{\vee}(\Delta kx)=e^{-ijK_{0}x}%
\phi^{\vee}(K_{0}x),\Delta k=K_{0},
\]
and%

\[
\supp\mathcal{F[}\phi_{j}^{\vee}(x)]\subset\lbrack\omega_{j-2}, \omega_{j+2}].
\]

From the basic property of B-spline functions, we know that
\[\sum_{j\in\mathbb{Z}}\phi_j(k) = \sum_{j\in\mathbb{Z}}B_4(\frac{k}{\Delta k}+2-j)\equiv 1,\quad \forall k\in\mathbb{R}.\]

Thus,
\begin{equation}\label{eq:sum1C4}
  \sum_{j=-m-1}^{m+2}\phi_j(k) = 1 ,\quad \forall k\in [-m\Delta k, m\Delta k].
\end{equation}
In Eqn. \eqref{eq:sum1C4}, we define $\phi_{-m-1} = B_4(k/\Delta k + m+3)$, $\phi_{m} = B_4(k/\Delta k - m+2)$ and $\phi_{m+1} = B_4(k/\Delta k -m +1 )$.

\bigskip Now, with the POU in (\ref{pou}), we can decompose the target
function $f(x)$ in the Fourier space as follows,%

\begin{equation}
\widehat{f}(k)=%
{\displaystyle\sum\limits_{j=-m}^{m-1}}
\phi_{j}(k)\widehat{f}(k)\triangleq%
{\displaystyle\sum\limits_{j=-m}^{m-1}}
\widehat{f_{j}}(k). \label{k-decomp}%
\end{equation}
%

Equation (\ref{k-decomp}) will give a corresponding decomposition in $x$-space%

\begin{equation}
f(x)=%
{\displaystyle\sum\limits_{j=0}^{m}}
f_{j}(x), \label{x-decomp}%
\end{equation}
where
\[
f_{j}(x)=\mathcal{F}^{-1}[\widehat{f_{j}}](x).
\]

The decomposition (\ref{x-decomp}) involves $m+1$ function $f_{j}(x)$ whose
frequency spectrum is limited to $[\omega_{j-1,}\omega_{j+1}],$ therefore, a
simple phase shift could translate its spectrum to $[-K_{0},K_{0}]$, then it
could be learned quickly by a DNN with $n_{0}$ training epoches.

\begin{itemize}
\item \bigskip A parallel phase shift DNN (PhaseDNN) algorithm
\end{itemize}

We define the first approximation to $f(x)$ by $T_{\star}(x,\theta^{(n_{0}%
)}),x\in R,$%

\begin{equation}
T_{\star}(x,\theta^{(n_{0})})=f_{\text{DNN}}(x)\approx f(x),\label{f1(x)}%
\end{equation}
based on given $N$-training data%

\begin{equation}
\{x_{i},y_{i}=f(x_{i})\}_{i=1}^{N}. \label{trainingdata}%
\end{equation}


\bigskip

Let us consider the residual between the function $f(x)$ and its first
DNN\ approximation $f_{\text{DNN}},$%
\begin{equation}
r(x)=f(x)-f_{\text{DNN}}(x).\label{R1}%
\end{equation}
We should have
\begin{equation}
\left\vert \mathcal{F}[r](k)\right\vert \ll1\text{ for }\left\vert
k\right\vert <K_{0},
\end{equation}
namely,
\begin{equation}
\mathcal{F}[f](k)\approx\mathcal{F}[f_{\text{DNN}}(x)](k),\quad k\in
(-K_{0},K_{0}).\label{eq:1psdnn}%
\end{equation}

Our next step is to learn this residual function $r(x)$ based on its training data%

\begin{equation}
\{x_{i},r_{i}=r(x_{i})\}_{i=1}^{N}. \label{r-trainingdata}%
\end{equation}

\bigskip In order to learn $r(x)$ for its frequency information for
$|k|>K_{0},$ we will apply the decomposition (\ref{x-decomp}) to $r(x)$%

\begin{equation}
r(x)=%
{\displaystyle\sum\limits_{j=0}^{m}}
r_{j}(x),
\end{equation}
where
\begin{equation}
r_{j}(x)=\mathcal{F}^{-1}[\widehat{r_{j}}(k)]=\mathcal{F}^{-1}[\phi
_{j}(k)\widehat{r}(k)](x).\label{rj(x)}%
\end{equation}
The training data for $\{r_{j}(x_{i})\}_{i=1}^{N}$ will be computed by
(\ref{rj(x)}), which can be implemented in the data $x$-space through the
following convolution%

\begin{align}
r_{j}(x_{i}) &  =\phi_{j}^{\vee}\ast r(x_{i})=\int_{-\infty}^{\infty}\phi
_{j}^{\vee}(x_{i}-s)r(s)ds\nonumber\\
&  \approx\frac{2\delta}{N_{s}}%
{\displaystyle\sum\limits_{x_{s}\in(x_{i}-\delta,x_{i}+\delta)}}
\phi_{j}^{\vee}(x_{i}-x_{s})r(x_{s}),\label{convol}%
\end{align}
where $\delta$ is chosen such that the kernel function
$\vert$%
$\phi_{j}^{\vee}(\delta)|=|\phi^{\vee}(K_{0}\delta)|$ is smaller than some
given error tolerance, $N_{s}$ is the number of data points inside
$(x_{i}-\delta,x_{i}+\delta).$

\begin{figure}[ptbh]
\center \includegraphics[scale=0.7]{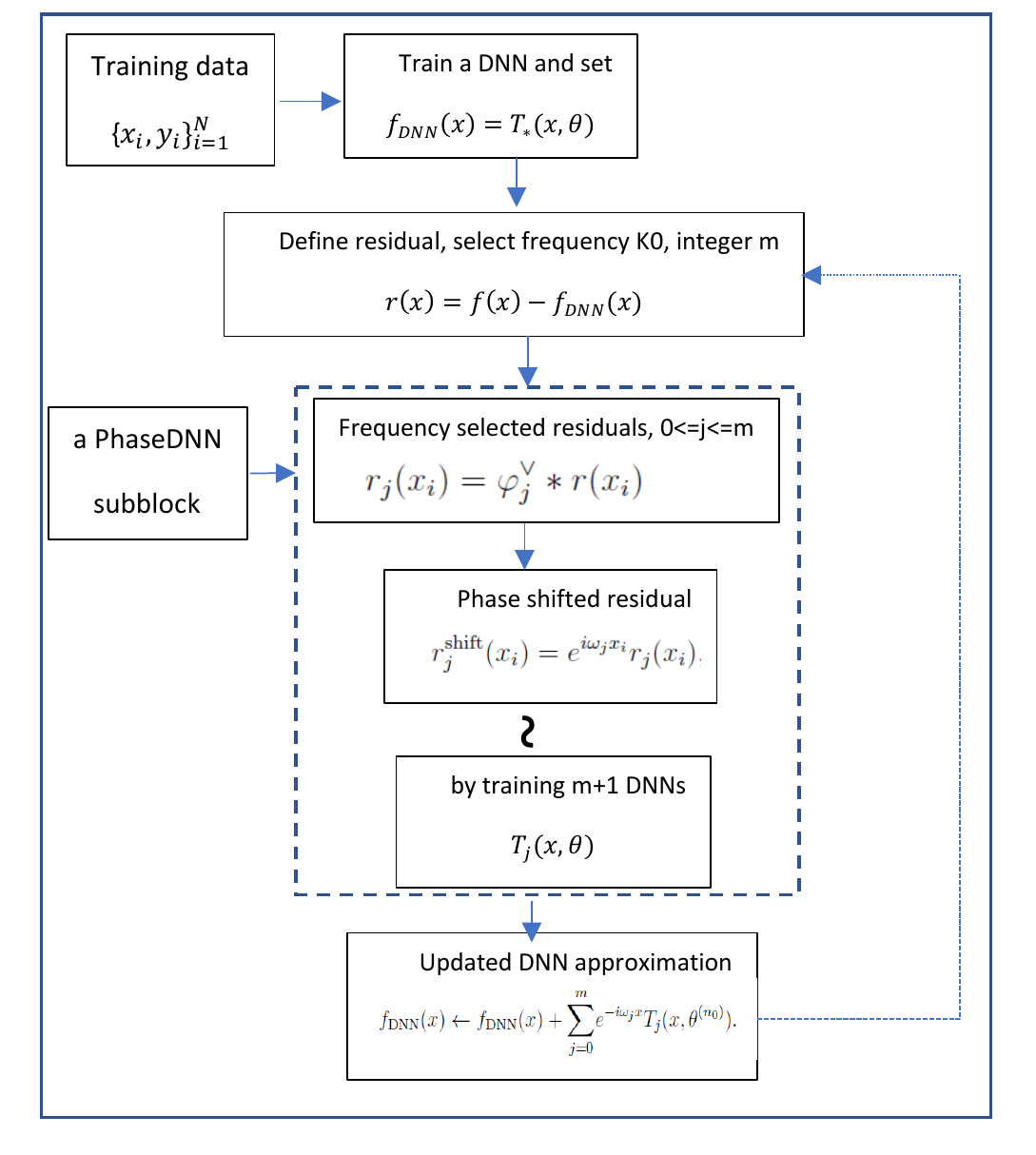}
\caption{Flowchart of Parallel PhaseDNN} with shifts for $\vec k \in R^d$ by an alternating coordinate direction sweep. %
\label{PhaseDNN}%
\end{figure}

Now as the support of $\widehat{r_{j}}(k)$ is $[\omega_{j-1,}\omega_{j+1}]$,
then $\widehat{r_{j}}(k-\omega_{j})$\ will have its frequency spectrum in
$[-\Delta k,\Delta k]$ $=[-K_{0},K_{0}],$ therefore its inverse Fourier
transform $\mathcal{F}^{-1}\left[  \widehat{r_{j}}(k-\omega_{j})\right]
,$denoted as $\ $%

\begin{equation}
r_{j}^{\text{shift}}(x)=\mathcal{F}^{-1}\left[  \widehat{r_{j}}(k-\omega
_{j})\right]  (x),\label{rjshift}%
\end{equation}
can be learned quickly by a DNN
\begin{equation}
T_{j}(x,\theta^{(n_{0})})\ \approx r_{j}^{\text{shift}}(x),0\leq j\leq m,
\end{equation}
in $n_{0}$-epoches of training. Moreover, we know that%

\begin{equation}
r_{j}^{\text{shift}}(x_{i})=e^{i\omega_{j}x_{i}}r_{j}(x_{i}),1\leq i\leq
N,\label{dataShift}%
\end{equation}
which provides the training data for $r_{j}^{\text{shift}}(x)$. Equation
(\ref{dataShift}) shows that once $r_{j}^{\text{shift}}(x)$ is learned,
$r_{j}(x)$ is also learned by removing the phase factor.$\mathcal{\ }$%
\begin{equation}
r_{j}(x)\approx e^{-i\omega_{j}x}T_{j}(x,\theta^{(n_{0})}).
\end{equation}
Now all $r_{j}(x)$ $0\leq j\leq m$ have been learned, we will have an
approximation to $r(x)$ as%

\begin{equation}
r(x)\approx%
{\displaystyle\sum\limits_{j=0}^{m}}
e^{-i\omega_{j}x}T_{j}(x,\theta^{(n_{0})}).
\end{equation}
Finally, we will obtain our Phase shift DNN (PhaseDNN) approximation of $f(x)$, as illustrated in the
flowchart in Fig. \ref{PhaseDNN}
over all frequency range $[-M,M]$ as follows%

\begin{equation}
f_{\text{DNN}}(x)\leftarrow f_{\text{DNN}}(x)+%
{\displaystyle\sum\limits_{j=0}^{m}}
e^{-i\omega_{j}x}T_{j}(x,\theta^{(n_{0})}). \label{DNNupdate}%
\end{equation}

\begin{Rem}
(\textbf{Parallelism of PhaseDNN}) It is clear the learning for each part of
residual function $r_{j}(x)$ can be done in parallel, the PhaseDNN is a
version of domain decomposition method in $k$-space.
\end{Rem}

\begin{Rem}
(\textbf{Recursive version of PhaseDNN}) After the updating step
(\ref{DNNupdate}), in principle, we can repeat the process by starting at
(\ref{R1}) with the new $f_{\text{DNN}}(x)$ to further reduce any remaining
error in all frequency range (see Fig. \ref{PhaseDNN}).
\end{Rem}

\begin{Rem}
(\textbf{Frequency} \textbf{Adaptivity of PhaseDNN}) The
frequency-selected residual $r_{j}(x_{i})$ requires the evaluation of the
convolution (\ref{convol}), which implicitly assumes sufficient data
information is available in the neighborhood of $x_{i}$. In case such an
assumption is not valid, we basically do not have enough data to extract the
frequency information for the target function. In all practical sense of
learning, we should assume the target function do not have the information of
this specific frequency, and there is no need to learn this function
$r_{j}(x)$ near $x_{i}$ and we could simply set $r_{j}(x)=0$ for $x\in
(x_{i}-\delta,x_{i}+\delta)$, which is what could happen for a constant (or
low) frequency function $r(x)$. As this process is done locally in space and
frequency, the PhaseDNN will be able to achieve frequency adaptivity
commonly associated with wavelet approximation of signals and images.

{$\bullet$ {\bf Adaptivity in frequency and local radial Basis fitting} } A pre-analysis of the training data could be done
to investigate the frequency content of the data, so only those frequencies will be considered
when phase shifts are used for error reduction. Firstly, we group the training data $x_i$ into
$N_c$ clusters (disjoint or overlapping). Secondly, in a sphere centered at the center of a cluster, a local radial
basis function fitting of training data $\{ x_i, y_i \}$ falling in the sphere
 will be carried out \cite{wendland04}. Finally, we apply an 1-D discrete Fourier transform over the fitting function along each
coordinate direction within the cluster to get the frequency content for the data in the cluster. Once this
analysis is done for each cluster, we will have a collection of frequencies and those with
significant magnitude will be used as candidates for the phase shifts in the PhaseDNN.

\end{Rem}

\begin{Rem}
(\textbf{Selection of $K_{0}$ and }$n_{0}$ and pre-processing training data)
In principle, any nonzero frequency threshold $K_{0}$ can be used in the
implementation of PhaseDNN, which will dictate the size of $n_{0}$ to achieve
sufficient convergence of the underlying DNN over the frequency range
$[-K_{0},K_{0}]$. Therefore, a careful analysis and numerical experiment of
the specific low frequency convergence of the DNN will be needed.

As $n_{0}$ is selected to train the initial network in PhaseDNN to learn low
frequent content of the data, a low-pass filtering on the training data could
be used to select appropriate size of $n_{0}$ needed for achieve convergence
within $K_{0}$ frequency.
\end{Rem}

\begin{Rem}
(\textbf{Avoiding curse of dimensionality}) It is clear the proposed
phase shift DNN can be extended to learn function $y=f(\overrightarrow
{x}),\overrightarrow{x}\in R^{d}$ by making frequency shift along various
$\overrightarrow{k}$ unit vector directions.  However,  for high dimension problems, the shifts employed in the frequency $\vec k \in R^d$-space
to learn $r(x)$ in (\ref{R1}) should be done
in an alternating direction sweep over the $\vec k$-coordinate directions to avoid the curse
of dimensionality. Namely, we will carry out $m+1$ shifts along $k_1$ direction, learn $m+1$ DNNs for the phase shifts only along $k_1$ direction, which will
be used to update $f_{DNN}(x)$ in (\ref{DNNupdate}). Then, we learn another $m+1$ DNNs for the shifts along $k_2$ direction, etc, until all directions
in $\vec k$ are touched. In all, $d(m+1)$ DNNs will have been trained for one sweep of all $\vec k$ coordinate directions at a cost $O(d)$.
This alternating direction sweep in $\vec k$ space can be repeated to improve the
accuracy.

Another way to improve the efficiency of the $\vec k$ alternating direction sweep is to introduce matching random rotation of the coordinate systems in both
$\vec x$ and $\vec k$ spaces, so the
rotated $\vec k$ coordinate directions will be able to detect the high frequency in non-uniform data during the sweep more effectively.

\end{Rem}

\section{Numerical Result}
In this section, we will present some preliminary numerical results to demonstrate the capability of PhaseDNN to learn high frequency content of target functions. In practice, we could sweep over all frequency ranges. For the test function we have some rough idea of the the range of frequencies in the data, only a few frequency intervals are selected for the phase shift. 

We choose a target function $f(x)$ defined in $[-\pi, \pi]$ by
\begin{equation}\label{eq:target1}
  f(x) = \begin{cases}
           10(\sin x + \sin 3x), & \mbox{if } x\in [-\pi,0] \\
           10(\sin 23x + \sin 137x + \sin 203x), & \mbox{if }x\in [0,\pi].
         \end{cases}
\end{equation}

\begin{figure}[htbp]
  \centering
  \includegraphics[scale = 0.7]{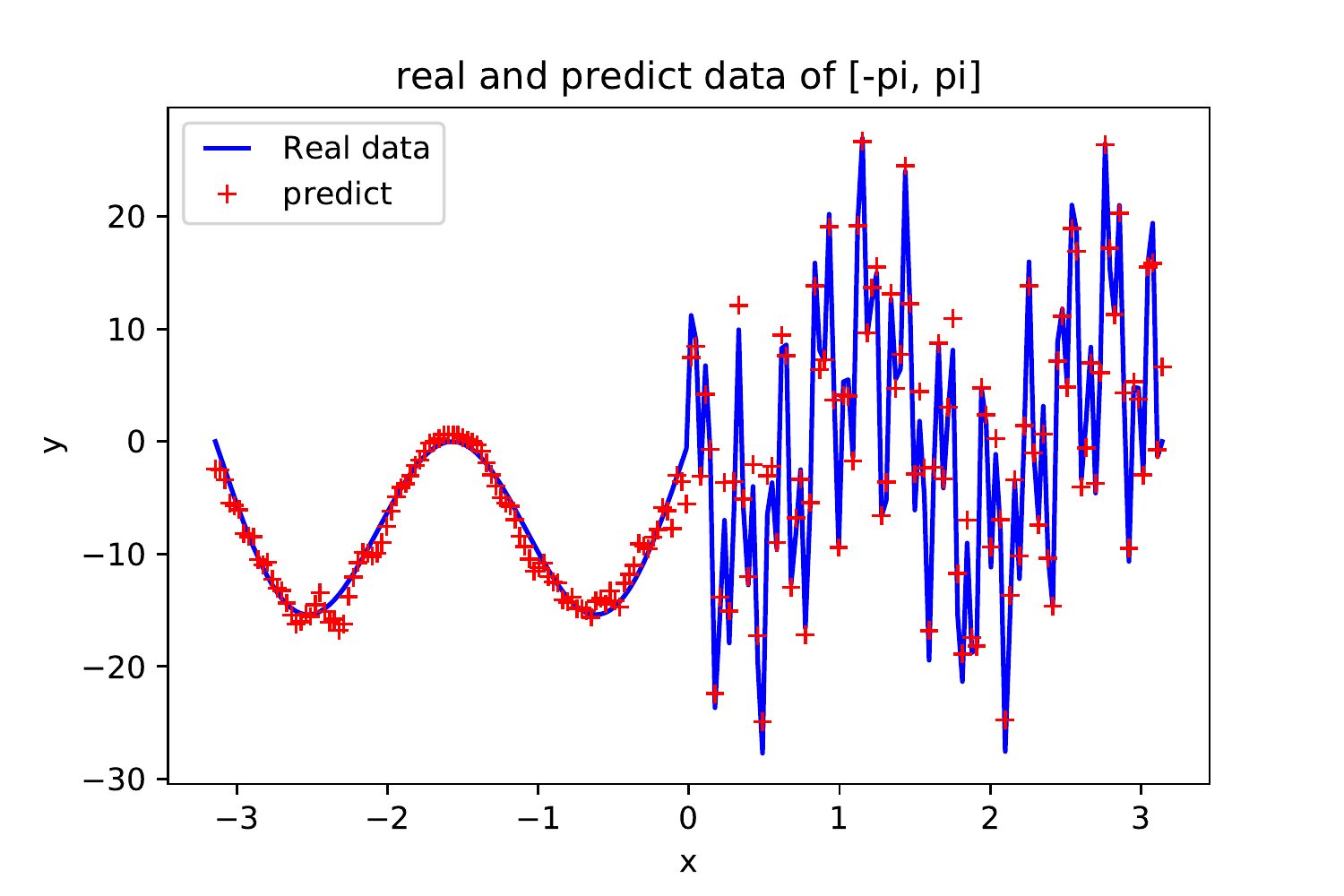}
  \caption{The testing result of $f_{DNN}(x)$ trained by phase DNN. The blue solid line is $f(x)$ and the data marked by $+$ in red is the value of $f_{DNN}(x)$ at testing data set.}\label{fig:fivefreq}
\end{figure}

Because the frequencies of this function is well separated, we need not to sweep all the frequencies in $[-\infty, +\infty]$. Instead,  we choose $\Delta k=5$, and use the following functions
\[
\begin{aligned}
&\phi_1(k) = \chi_{[-205,-200]}(k)  & \phi_2(k) = \chi_{[-140,-135]}(k)\\
& \phi_3(k) = \chi_{[-25,-20]}(k)  & \phi_4(k) = \chi_{[-5,0]}(k)\\
&\phi_5(k) = \chi_{[0,5]}(k)   & \phi_6(k) = \chi_{[20,25]}(k)\\
& \phi_7(k) = \chi_{[135,140]}(k)  & \phi_8(k) = \chi_{[200,205]}(k)
\end{aligned}
\]
to collect the frequency information in the corresponding frequency intervals and shift the center of the interval to the origin by a phase factor. For each $f_j(x) = \mathcal{F}^{-1}[\widehat{f}\phi_j](x)$, we construct two DNNs to learn its real part and imaginary part, separately. Every DNN have 4 hidden layers and each layer has 40 neurons. Namely, the DNN has the structure 1-40-40-40-40-1. The training data is obtained by 10,000 samples from the uniform distribution on $[-\pi,\pi]$ and the testing data is 500 uniformly distributed samples. We train these DNNs by Adam optimizer with training rate 0.0002 with 10 epochs of training for each DNN. The result is shown in Fig. \ref{fig:fivefreq}.

\begin{figure}
  \centering
  \scalebox{0.5}{\includegraphics[angle=90]{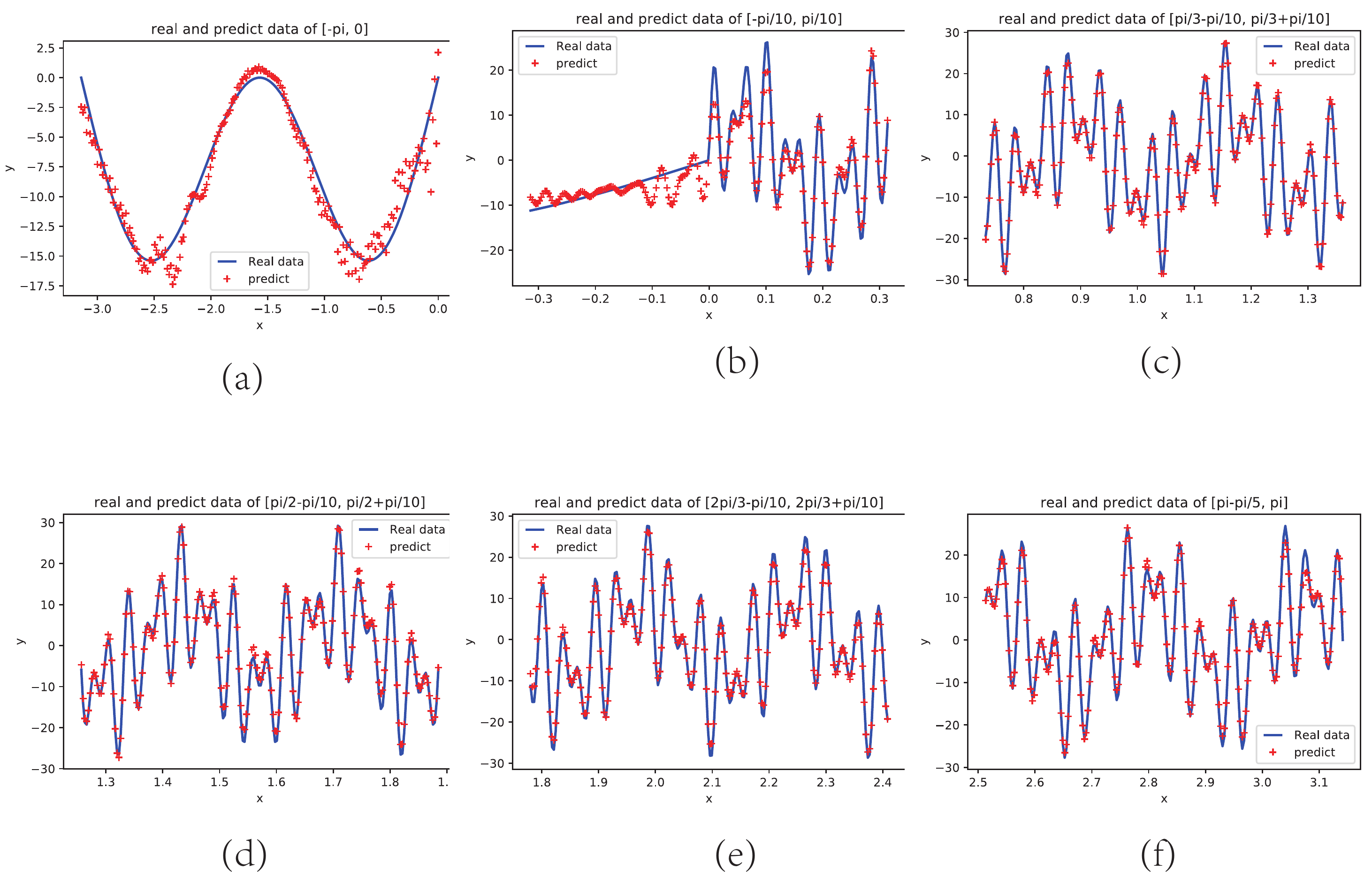}}
  \caption{The detail result of training in different intervals. The subfigure (a)-(f) shows the results in interval $[-\pi,0]$, $[-\pi/10,\pi/10]$, $[\pi/3-\pi/10,\pi/3+\pi/10]$, $[\pi/2-\pi/10,\pi/2+\pi/10]$, $[2\pi/3-\pi/10,2\pi/3+\pi/10]$ and $[\pi-\pi/10,\pi]$ correspondingly. The blue solid line is $f(x)$ and the data marked by $+$ in red is the value of $f_{DNN}(x)$ at testing data set.}\label{fig:fivefreqdetail}
\end{figure}

The detail of this training is shown in Fig. \ref{fig:fivefreqdetail}.

These figures clearly shows that phase DNN can capture the various high frequencies, from low frequency $\pm 1$, $\pm 3$ to high frequency $\pm 203$ quite well. The training error and time of phase DNN are collected in table \ref{tab:phaseDNN}

\begin{table}[htbp]
  \centering
  \begin{tabular}{|c|c|c|c|c|}
    \hline
    \,               & Convolution  & Training & Training  & Test\\
    \,               & time(s) & Time(s) &  Error & Error\\
    $[-205,-200]$    & 34.43 & 169.82 &  -  & - \\
    $[-140,-135]$    & 33.74 & 185.40 &  -  & - \\
    $[-25,-20]$      & 33.22 & 199.48 &  -  & - \\
    $[-5,0]$         & 9.70  & 214.39 &  -  & - \\
    $[0,5]$          & 9.78  & 230.46 &  -  & - \\
    $[20,25]$        & 33.21 & 247.34 &  -  & - \\
    $[135,140]$      & 33.38 & 263.89 &  -  & - \\
    $[200,205]$      & 32.68 & 279.39 &  -  & - \\
    Total            & 220.14& 1790.17& 0.01508 & 0.06831 \\
    \hline
  \end{tabular}
  \caption{The training time and error statistics. For each $j$, the training time is the sum of training time of real and imaginary part. For each DNN, epoch is chosen to be 10.}\label{tab:phaseDNN}
\end{table}
It is shown that taking the advantage of vectorization, the convolution step costs about 20\% time of training. Because in different interval, $f_j(x)$ can be trained in parallel, PhaseDNN is ideal to take advantage of parallel computing architectures.

In comparison with one single DNN, we have used a 24 hidden layers with 80 neurons per hidden layer with the same amount of training data, the loss is still around 100 after more than 5000 epoch of training taking over 22 hours nonstop running on the same workstation.

\section{Discussion and Conclusion}

In this paper, we proposed a parallel PhaseDNN, consisting of a series of
commonly-used DNNs in parallel, to achieve uniform frequency convergence of
deep neural network in approximating a function. In contrast to the wavelet
multi-resolution approximation with its higher resolution wavelet subspaces
generated by dilation of mother wavelet to provide approximation to higher
frequency components, the PhaseDNN uses a phase shift strategy to handle
higher frequency, which allows us to make use of the fast convergence of
common DNN in the low frequency range. The PhaseDNN employs original training
data, which are easily modified (i.e., a multiplication by a phase factor) to
carry out the required phase shift operation.

\bigskip Two additional comments on the PhaseDNN are given below.

\begin{itemize}
\item \textbf{Training data and PhaseDNN}. The convolution computation
(\ref{convol})\ required in obtaining the training data for the selected
frequency component of the residual $r_{j}(x)$ implicitly implies that the
training data location $x$ are dense enough to extract the information of the
given frequency range$.\ $Similarly, The frequency shift in the PhaseDNN in
(\ref{dataShift}) also implicitly requires that the training data location $x$
are dense enough to be uni-solvent to represent the Fouirer mode $e^{\pm
i\omega_{j}x}.\ $Otherwise due to the alias phenomenon of the Fourier modes on
a discrete set, the simple multiplication by $e^{\pm i\omega_{j}x}$ on the
training data will not achieve the desired frequency shift.

\item \textbf{Spatial} \textbf{and} \textbf{Frequency adaptive approximation}.
The frequency shifts can be done in any order, therefore, for problems for
prior knowledge of frequency information or need of knowledge for specific
frequency such as segmentation, the PhaseDNN can be used to obtain convergence
in the desired frequency ranges. As mentioned in Remark 3, each DNN in the
PhaseDNN is linked to the local spatial and frequency, therefore PhaseDNN can
achieve also spatial adaptivity in the construction of the its DNNs.
\end{itemize}

\bigskip

In future work, more numerical tests will be carried out to understand
the potential and behavior of the proposed PhaseDNN and calibrate the
efficiency and accuracy of the PhaseDNN over a stand-alone DNN, especially for high dimensional problems.

\section*{Acknowledgment}

Authors thank Dr. Zhiqin Xu for useful discussion on the F-principle of DNNs.

\end{document}